\documentclass[preprint, aps, pra, showpacs]{revtex4-1}
\usepackage{eurosym}
\usepackage{graphicx}
\usepackage{dcolumn}
\usepackage{bm}
\usepackage{amsmath}
\usepackage{color}

\begin{document}

\title{Intra- and intercycle interference of electron emission in laser assisted XUV atomic ionization}

\author{A. A. Gramajo}
\affiliation{Centro At\'omico Bariloche (CNEA) and CONICET, 8400 Bariloche, Argentina}

\author{R. Della Picca}
\affiliation{Centro At\'omico Bariloche (CNEA) and CONICET, 8400 Bariloche, Argentina}

\author{C. R. Garibotti}
\affiliation{Centro At\'omico Bariloche (CNEA) and CONICET, 8400 Bariloche, Argentina}

\author{D. G. Arb\'o}

\affiliation{Institute for Astronomy and Space Physics IAFE
(CONICET-UBA), CC 67, Suc. 28, C1428ZAA, Buenos
Aires, Argentina}

\date{\today}

\begin{abstract}
We study the ionization of atomic hydrogen in the direction of polarization due to a linearly polarized XUV pulse in the presence a strong field IR. We describe the photoelectron spectra as an interference problem in the time domain. Electron trajectories steming from different optical laser cycles give rise to \textit{intercycle} interference energy peaks known as sidebands. These sidebands are modulated by a grosser structure coming from the \textit{intracycle} interference of the two electron trajectories born during the same optical cycle. We make use of a simple semiclassical model which offers the possibility to establish a connection between emission times and the photoelectron kinetic energy. We compare the semiclassical predictions with the continuum-distorted wave strong field approximation and the \textit{ab initio} solution of the time dependent Schr\"odinger equation. We analyze such interference pattern as a function of the time delay between the IR and XUV pulse and also as a function of the laser intensity.
\end{abstract}

\maketitle

\preprint{APS/123-QED} 


\section{\label{sec:level1}Introduction}

New sources of coherent XUV and soft-X-ray radiations delivering pulses with
durations in the femtosecond range and with unprecedented high intensities
open new perspectives in atomic and molecular physics. Such sources produced
from either high-order harmonics or from X-ray free-electron laser (XFEL)
paves the way to explore the dynamics of atomic, molecular, and even
solid-surface systems undergoing inner-shell transitions. In this way,
multi-photon spectroscopy involving synchronized IR and XUV pulses in the
strong field regime can be achieved. The photoelectron spectra from rare gas
atoms have been extensively studied in the simultaneous presence of two
pulses from the XUV source and from an IR laser with a time-controlled delay
working as a pump-probe experiment \cite{Taieb96,Schins96}.

The two-color  multiphoton ionization where one of the two radiation fields has low intensity and relatively high frequency  while the other is intense with a low frequency is usually known as laser assisted photoemission (LAPE). Depending on the features of both laser fields (typically the pulse durations), two well-known regimes --streak camera and sideband-- can be distinguished \cite{Itatani02,Fruehling09,Wickenhauser06,Nagele11,Duesterer13,Kazansky10,Kazansky09a}.
In the former, the XUV pulse is much shorter than the IR period $T_L=2\pi/\omega_L$ and, therefore, the electron behaves like a classical particle that gets linear momentum from the IR laser field at the instant of ionization \cite{Drescher05}.
On the other hand, when the XUV pulse is longer than the laser period $T_{L}$,
the photoelectron energy spectrum shows a main line
associated with the absorption of one XUV photon accompanied by sideband
lines, located more or less symmetrically on its sides. The equally spaced
sidebands, that are separated from each other by $\hbar \omega_{L}$, are
associated with additional exchange of laser photons of frequency 
$\omega_{L}$ through absorption and stimulated emission processes. The
analysis of the resulting two color photoelectron spectra can provide
information about the high-frequency pulse duration, laser intensity, and
the time delay between the two pulses. However, the intermediate situation 
the duration of the XUV pulse is comparable to the laser period has not been
thoroughly studied.

An accurate theoretical description of the process must be based on quantum
mechanical concepts, i.e. by solving \textit{ab initio} the time dependent
Sch\"{o}dinger equation (TDSE) for the atomic system in the presence of the
two pulses. However, the precise calculation of the response of a rare gas
atom presents considerable difficulties. The numerical resolution of the
TDSE for a multi-electron system rely on the single-active electron
approximation, with model potentials that permit one to reproduce the bound
state spectrum of the atom with a satisfactory accuracy 
\cite{Nandor99,Muller99}, but results are sensitive to the used approximation.
Simplified models have been proposed, such as the
the Simpleman's classical model \cite{Schins94}, the soft-photon approximation 
\cite{Maquet07,Gimenez-Galan13,Taieb08,Kroll73} for large pulse durations case, 
and the strong field approximation (SFA) and Coulomb-Volkov approximation (CVA) in the streaking to sideband transition regime \cite{Kazansky10, Bivona10}. 
These models provide a useful description of some
general features. At the time of discussing the physical content of the
experimental data or full numerical results, it is instructive to compare
them to the qualitative predictions of a simplified analysis. 
In Ref. \cite{Arbo10a,Arbo10b}, starting from a semiclassical description of above-threshold
ionization (ATI) by a one color laser it has been identified the interplay
of \textit{intracycle} and \textit{intercycle} interferences between
trajectories of electrons emitted at different times, giving rise to a
description of the photoelectron spectra of direct electrons
as the interplay of such inter- and intracycle interference pattern.

In this paper we use a semiclassical approximation \cite{Arbo10a,Arbo10b}
to analyze the laser assisted
electron photoemission spectra of hydrogen atoms by a XUV pulse, 
particularly in the intermediate case where $\tau_{X}\sim T_L$ or few IR cycles. 
We show that the role of the IR laser field is threefold: (a) due to the average
wiggling of the electron it shifts the energy of the continuum states of the
atom by the ponderomotive energy $U_{p}$, (b) besides the absorption of the
high frequency photon, several IR photons can be absorbed or emitted in the
course of the ionization process, and (c) it is responsible for modulations in
the photoelectron energy spectrum. For (b), we show that the exchange of IR photons
in the energy domain can be interpreted as the interference
among different electron trajectories emitted by the atom at different
optical cycles giving origin to the formation of sidebands. More importantly,
for (c), the interfering electron trajectories within the same optical cycle
give rise to a well-determined modulation pattern encoding information of
the ionization process in the subfemtosecond time scale. 

The paper is organized as follows. In Sec. \ref{sec:level2} we describe the different
methods of calculating the photoelectron spectra for the case of laser
assisted XUV ionization: By solving the TDSE \textit{ab initio}, making use of the theory
of the strong field approximation (SFA), and a semiclassical model which
gives rise to simple analytical expressions. In Sec. \ref{results}, we present the
results and discuss over the comparison of results calculated within the
different methods. Concluding remarks are presented in Sec. \ref{conc}.
Atomic units are used throughout the paper, except when otherwise stated.


\section{\label{sec:level2}Theory and methods of laser-assisted photoemission}


We want to solve the problem of atomic ionization by an XUV pulse in the
presence of an IR laser both linearly polarized along the $\hat{z}$
direction. The time-dependent Schr\"{o}dinger equation (TDSE) in the single
active electron (SAE) approximation reads%
\begin{equation}
i\frac{\partial }{\partial t}\left\vert \psi (t)\right\rangle =H\left\vert
\psi (t)\right\rangle ,  \label{TDSE}
\end{equation}%
where the hamiltonian of the system within the dipole approximation in the
length gauge is expressed as%
\begin{equation}
H=\frac{\vec{p}^{2}}{2}+V(r)+\vec{r}\cdot \vec{F}_{X}(t)+\vec{r}\cdot \vec{F}%
_{L}(t).  \label{hamiltonian}
\end{equation}%
The first term in Eq. (\ref{hamiltonian}) corresponds to the active electron
kinetic energy, the second term is the potential energy of the active
electron due to the Coulomb interaction with the core, and the last two
terms correspond to the interaction of the atom with the electric fields $%
\vec{F}_{X}(t)$ and $\vec{F}_{L}(t)$ of the XUV pulse and IR laser,
respectively.

As a consequence of the interaction, the bound electron in the initial state 
$|\phi _{i}\rangle $ is emitted with momentum $\vec{k}$ and energy $%
E=k^{2}/2 $ into the final unperturbed state $|\phi _{f}\rangle $. The
photoelectron momentum distributions can be calculated as 
\begin{equation}
\frac{dP}{d\vec{k}}\mathbf{=}\left\vert T_{if}\right\vert ^{2},  \label{P}
\end{equation}%
where $T_{if}$ is the T-matrix element corresponding to the transition $\phi
_{i}\rightarrow \phi _{f}$.

\subsection{Time-Dependent Schr\"{o}dinger Equation}

The evolution of the electronic state $\left\vert \psi (t)\right\rangle $ is
governed by the TDSE in Eq. (\ref{TDSE}) for the Hamiltonian of Eq. (\ref%
{hamiltonian}). In order to numerically solve the TDSE in the dipole
approximation for the SAE, we employ the generalized
pseudo-spectral method \cite{Tong97,Tong00,Tong05}. This method combines the
discretization of the radial coordinate optimized for the Coulomb
singularity with quadrature methods to allow stable long-time evolution
using a split-operator representation of the time-evolution operator. Both
the bound as well as the unbound parts of the wave function $|\psi
(t)\rangle $ can be accurately represented. Due to the cylindrical symmetry
of the system the magnetic quantum number $m$ is conserved. After the end of
the laser pulse the wave function is projected on eigenstates $|k,\ell
\rangle $ of the free atomic Hamiltonian with positive eigenenergy $%
E=k^{2}/2 $ and orbital quantum number $\ell $ to determine the transition
amplitude $T_{if}$ to reach the final state $|\phi _{f}\rangle $ (see Refs.~%
\cite{Schoeller86,Messiah65,Dionis97}): 
\begin{equation}
T_{if}=\frac{1}{\sqrt{4\pi k}}\sum_{\ell }e^{i\delta _{\ell }(p)}\ \sqrt{2l+1%
}P_{\ell }(\cos \theta )\left\langle p,\ell \right. \left\vert \psi
(t_{f})\right\rangle .  \label{coulomb}
\end{equation}%
In Eq.~(\ref{coulomb}), $\delta _{\ell }(p)$ is the momentum-dependent
atomic phase shift, $\theta $ is the angle between the electron momentum $%
\vec{k}$ and the polarization direction $\hat{z}$, and
$P_{\ell }$ is the Legendre polynomial of degree $\ell $. In
order to avoid unphysical reflections of the wave function at the boundary
of the system, the length of the computing box was chosen to be 1200 a.u.\ ($%
\sim 65$ nm) which is much larger than the maximum quiver amplitude $\alpha
=F_{L0}/\omega _{L}^{2}=8$ a.u. at the intensity of $1.5\times 10^{13}$ W/cm$%
^{2}$ and the wavelength of 750 nm. The maximum angular momentum included
was $\ell _{\max }=200$.

\subsection{Strong Field Approximation}

Within the time-dependent distorted wave theory, the transition amplitude in
the prior form and length gauge is expressed as 
\begin{equation}
T_{if}=-i\int_{-\infty }^{+\infty }dt\left\langle \chi _{f}^{-}(\vec{r}%
,t)\right\vert \left[ \vec{r}\cdot \vec{F}_{X}(t)+\vec{r}\cdot \vec{F}_{L}(t)%
\right] \left\vert \phi _{i}(\vec{r},t)\right\rangle \,  \label{Tif}
\end{equation}%
where $\phi _{i}(\vec{r},t)=\varphi _{i}(\vec{r})\,\mathrm{e}^{\mathrm{i}%
I_{p}t}$ is the initial atomic state with ionization potential $I_{p}$ and $%
\chi _{f}^{-}(\vec{r},t)$ is the distorted final state. As the SFA
neglects the Coulomb distortion in the final channel, the distorted final
wave function can be written as $\chi_{f}^{-}(\vec{r},t)=\chi ^{V}(\vec{r},t),\,$
where \cite{Volkov} 
\begin{equation}
\chi ^{V}(\vec{r},t)=\,\frac{\exp {[\mathrm{i}}\left( \vec{k}+\vec{A}{(t)}%
\right) \cdot \vec{r}{]}}{(2\pi )^{3/2}}\exp {\left[ \frac{i}{2}%
\int_{t}^{\infty }dt^{\prime }\left( \vec{k}+\vec{A}(t^{\prime })\right) ^{2}%
\right] }\,  \label{Volkov}
\end{equation}%
is the length-gauge Volkov state and $\vec{A}{(t)}$ is the vector potential due
to the combined electron field%
\begin{equation}
\vec{F}(t)=\vec{F}_{L}(t)+\vec{F}_{X}(t).  \label{total-field}
\end{equation}%
For the sake of simplicity, hereinafter we consider ionization of a
hydrogenic atom of nuclear charge $Z=1.$

\subsection{Semiclassical model}

From TDSE and SFA calculations we have observed that the first and second
terms in Eq. (\ref{Tif}) are well separated in the energy domain: Whereas
the single-photon XUV ionization (first term) leads to ionization of
electrons with final kinetic energy close to $E\simeq \hbar \omega
_{X}-I_{p} $ (with $\omega _{X}$ the photon energy of the XUV pulse), the
ionization due to the IR laser (second term) leads to ionization of
electrons with final kinetic mostly less than twice its ponderomotive energy 
$E\lesssim 2U_{p}.$ If we focus on the emission due to the XUV pulse around
energy $E\simeq \omega _{X}-I_{p},$ the contribution of the second term in
Eq. (\ref{Tif}) is negligible wether $U_p \ll \omega_X - I_p$.
Therefore, inserting Eq. (\ref{Volkov}) into
Eq. (\ref{Tif}), the transition amplitude within the SFA reads%
\begin{equation}
T_{if}=-i\int_{-\infty }^{+\infty }dt\ \vec{d}
\big(\vec{k}+\vec{A}%
(t)\big)\cdot \vec{F}_{X}(t)
\exp \left\{ -{i\int_{t}^{\infty }dt^{\prime }\left[ \frac{\left( \vec{k}%
+\vec{A}(t^{\prime })\right) ^{2}}{2}+I_{p}\right] }\right\} \,.
\label{Tifx}
\end{equation}%
where the dipole element 
$\vec{d}(\vec{v})$ is given by 
\begin{equation}
\vec{d}
(\vec{v})=\frac{1}{(2\pi )^{3/2}}\int d\vec{r}\exp {[-}i\vec{v%
}\cdot \vec{r}{]\ }\vec{r}\ \varphi _{i}(\vec{r})
\label{dipole}.
\end{equation}

Let us suppose that the XUV pulse has the form $\vec{F}_{X}(t)=\hat{z}%
F_{X0}(t)\cos \omega _{X}t$ where $F_{X0}(t)$ is a slowly nonzero varying envelope
function in the time interval with duration $\tau _{X}$. 
In this case, writing $\cos \omega_X t=\left[ \exp (i\omega _{X}t)+\exp
(-i\omega _{X}t)\right] /2$, the transition amplitude can be written as $%
T_{if}=T_{if}^{+}+T_{if}^{-},$ where $T_{if}^{+}$ and $T_{if}^{-}$
correspond to the absorption and emission of an XUV photon, respectively. We
can discard the emission term since it does not lead to ionization. In other
words, according to the rotating wave approximation $T_{if}^{-}$
contribution lays in an energy domain close to $E\simeq -\omega _{X}-I_{p}$
which is not in the continuum. Therefore, we can write 
\begin{equation}
T_{if}=T_{if}^{+}=-\frac{i}{2}\int_{-\infty }^{+\infty }dt\ d_z
\left( \vec{k}+\vec{A}(t)\right) F_{X0}(t)\, \exp {\left[ iS(t)\right]} ,
\label{Tifg}
\end{equation}%
where%
\begin{equation}
S(t)=-\int_{t}^{\infty }dt^{\prime }\left[ \frac{\left( \vec{k}+\vec{A}%
(t^{\prime })\right) ^{2}}{2}+I_{p}-\omega _{X}\right]  \label{action}
\end{equation}%
is the generalized action for the case of LAPE for absorption of a single
XUV photon. As the frequency of the XUV pulse is much higher than the IR
laser frequency, for XUV pulses not much more intense than the IR laser, we
can consider the vector potential as due to the laser field only, i.e., $%
\vec{A}(t)\simeq \vec{A}_{L}(t),$ neglecting its XUV contribution \cite{DellaPicca13}.

In order to calculate the transition amplitude we need to solve the
four-dimensional integral of equations (\ref{Tifg} and \ref{dipole}). When the XUV pulse in shorter
than the period of the IR laser, i.e., $\tau _{X}<T_{L}=2\pi /\omega _{L},$
the electron is emitted with kinetic energy that depends of the vector
potential at the ionization time, what is known as streak camera \cite%
{Drescher05,Goulielmakis04,Itatani02,Meyer06,Kazansky10}. However, from now
on, we restrict to the case where the XUV pulse is comparable to or longer
than the period of the IR laser, i.e., $\tau _{X}\gtrsim T_{L}.$
Specifically, the SCM consists of solving the time integral of Eq. (\ref%
{Tifg}) by means of the saddle point approximation (SPA) \cite%
{Chirila05,Corkum94,Ivanov95,Lewenstein95}. In this sense, the transition
probability can be written as a coherent superposition of classical
trajectories with the same final momentum $\vec{k}$ as%
\begin{equation}
T_{if}=\sum_{t_{s}}\frac{\sqrt{2\pi }F_{X0}(t_{s})d_{z}
(\vec{k}+\vec{%
A}{(t}_{s}{)})}{\left\vert \left[ \vec{k}+\vec{A}{(t}_{s}{)}\right] \cdot 
\vec{F}_{L}(t_{s})\right\vert ^{1/2}}\exp \left[ iS(t_{s})\right] ,
\label{Tifsaddle}
\end{equation}%
where $t_{s}$ are the ionization times corresponding to the stationary
points of the action, i.e., $dS(t_{s})/dt=0.$ Then, from Eq. (\ref{action}),
the ionization times fulfill the equation%
\begin{equation}
\frac{\left( \vec{k}+\vec{A}(t_{s})\right) ^{2}}{2}+I_{p}-\omega _{X}=0.
\label{saddle}
\end{equation}

\begin{figure}[tbp]
\includegraphics[width=0.5\textwidth]{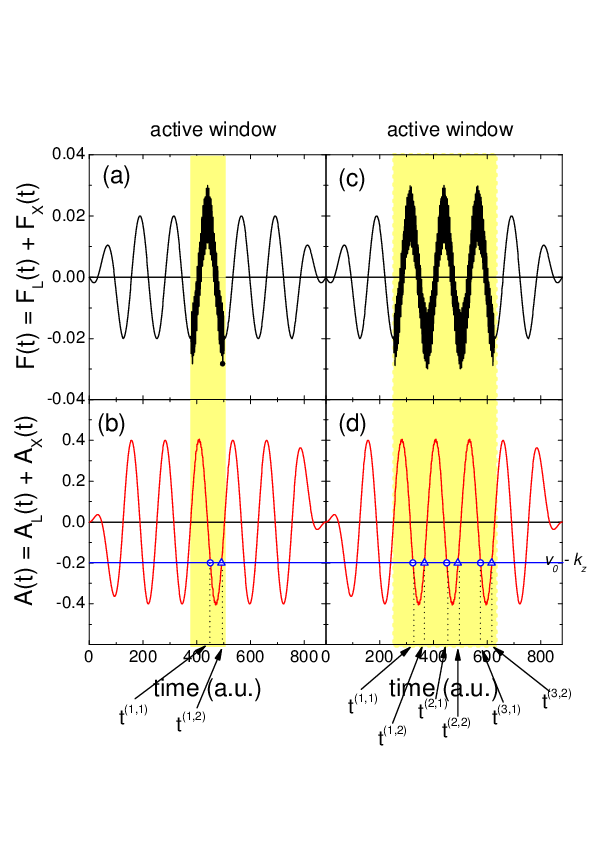}
\caption{(Color online) Total electric field $%
F(t)=F_{L}(t)+F_{X}(t)$ [(a) and (c)] and vector potential $%
A(t)=A_{L}(t)+A_{X}(t)$ [(b) and (d)] as a function of time. The IR laser
parameters are $F_{L0}=0.02,$ $\omega _{L}=0.05,$ $\tau _{L}=7T_{L},$ and
the XUV pulse with parameters $F_{X0}=0.01$ and $\omega _{X}=1.5.$ The
duration of the XUV pulse is $\tau _{X}=T_{L}$ in (a) and (b), and $\tau
_{X}=3T_{L}$ in (c) and (d). In (b) and (d) the electron emission early
(late) times for a given final momentum $k_{z}$ are marked by circles
(triangles).}
\label{fields}
\end{figure}

Let us consider an IR electric field $\vec{F}_{L}(t)=F_{L0}\cos
\omega _{L}t\ \hat{z}$ which is a good approximation for long laser pulses
where we can neglect the effect of the envelope. The vector potential is,
thus, $\vec{A}(t)=-(F_{L0}/\omega _{L})\sin \omega _{L}t\ \hat{z}$. 
In the
following we restrict our analysis to forward emission in the direction of
polarization, i.e., $k_z \geq 0$ and $k_{\rho }=0$.
Under $F_{L0}/\omega_L < v_0$ condition, where
$v_{0}=\sqrt{2\left( \omega _{X}-I_{p}\right) }$ is the
electron momentum for ionization of an XUV pulse only, there are two ionization times per
optical cycle. They are the early ionization time $t^{(j,1)}$ and the late ionization
time $t^{(j,2)}$ corresponding to the $j-$th optical cycle, with $%
t^{(j,\alpha )}=t^{(1,\alpha )}+2\pi /\omega _{L}(j-1)$ with $\alpha =1,2$
[see Fig. \ref{fields} (b) and (d)]$.$ In order to find the expressions for $%
t^{(j,\alpha )},$ we must consider two cases: $k_{z}\geq v_{0}$ and $%
k_{z}<v_{0}$, with solutions%
\begin{eqnarray}
t^{(1,1)} &=&\frac{1}{\omega _{L}}\sin ^{-1}\left[ \frac{\omega _{L}}{F_{L0}}%
\left( k_{z}-v_{0}\right) \right] ,  \label{ioniztimes+} \\
t^{(1,2)} &=&\frac{\pi }{\omega _{L}}-t^{(1,1)},  \nonumber
\end{eqnarray}%
and 
\begin{eqnarray}
t^{(1,1)} &=&\frac{-1}{\omega _{L}}\sin ^{-1}\left[ \frac{\omega _{L}}{F_{L0}%
}\left( k_{z}-v_{0}\right) \right] +\frac{\pi }{\omega _{L}},  \nonumber \\
t^{(1,2)} &=&\frac{3\pi }{\omega _{L}}-t^{(1,1)},  \label{ioniztimes-}
\end{eqnarray}%
respectively.

Real ionization times are in the framework of classical trajectories of
escaping electrons. Eq. (\ref{saddle}) delimits the classical realm to
momentum values $\left\vert \frac{\omega _{L}}{F_{L0}}\left(
k_{z}-v_{0}\right) \right\vert <1.$ In other words, the possible classical
values of the electron momentum along the positive polarization axis
are restricted to $v_{0}-F_{L0}/\omega_{L}\leq k_{z}\leq v_{0}+F_{L0}/\omega _{L}$.
Outside this domain,
ionization times are complex due to the non-classical nature of such
electron trajectories. The imaginary part of these ionization times gives
rise to exponentially decaying factors, for what complex (non-classical)
trajectories posses minor relevance compared to real (classical) ones. In
consequence, hereinafter we restrict our SCM to classical trajectories.

Including Eq. (\ref{Tifsaddle}) into Eq. (\ref{P}), The ionization
probability is calculated as%
\begin{equation}
\left\vert T_{if}\right\vert ^{2}=\left\vert \sum_{\alpha ,j}\frac{\sqrt{%
2\pi }F_{X0}(t^{(j,\alpha )})d_z \big(k_{z}+A{(t^{(j,\alpha )})}\big)}{%
\left\vert \left[ k_{z}+A{(t^{(j,\alpha )})}\right] F_{L}(t^{(j,\alpha
)})\right\vert ^{1/2}}\exp \left[ iS(t^{(j,\alpha )})\right] \right\vert
^{2},  \label{bsol}
\end{equation}%
with $\alpha =1(2)$ corresponding to the early (late) release times of Eq. (%
\ref{ioniztimes+}) [Eq. (\ref{ioniztimes-})]. Assuming now that the
depletion of the ground-state is negligible, the ionization rate [the
prefactor before the exponential in Eq. (\ref{bsol})] is identical for all
subsequent ionization bursts (or trajectories) and is only a function of the
time-independent final momentum $k_{z}$. This is only valid for the special
case that the IR laser is a plane wave with no envelope and also the
envelope of the XUV pulse $F_{X0}(t)$ is time independent, i.e., flattop
pulse, and where the effect of the Coulomb potential on the receding
electron is negliglible (SFA). We consider that the flattop XUV pulse
comprises an integer number of IR optical cycles, i.e., $\tau
_{X}=NT_{L}=2N\pi /\omega _{L}$. As there are two interfering trajectories
per optical cycle of the IR field, the total number of interfering
trajectories with final momentum $k_{z}$ is $M=2N$, with $N$ being the
number of IR cycles involved. The sum over interfering trajectories in
Eq.\ (\ref{bsol}) can thus be decomposed into those associated with the 
two release times within the same cycle and those associated with release
times in different cycles. Consequently, the momentum distribution
[Eq. (\ref{bsol})] can be written within the SCM as 
\begin{equation}
\left\vert T_{if}\right\vert ^{2}=\Gamma (k_{z})\left\vert
\sum_{j=1}^{N}\,\sum_{\alpha =1}^{2}e^{iS(t^{(j,\alpha )})}\right\vert ^{2},
\label{interf}
\end{equation}%
where the second factor on the right hand side of Eq. (\ref{interf})
describes the interference of the $2N$ classical trajectories with final
momentum $k_{z}$, and $t^{(j,\alpha )}$ is a function of $k_{z}$ through
equations (\ref{ioniztimes+}) and (\ref{ioniztimes-}) wether $k_{z}\geq
v_{0} $ and $k_{z}<v_{0}$, respectively. The ionization probability
$\Gamma (k_{z})$ is given by%
\begin{equation}
\Gamma (k_{z})=2\pi \frac{\left\vert F_{X0}(t^{(j,\alpha )})\right\vert
^{2}\left\vert d_{z}
\big(k_{z}+A{(t^{(j,\alpha )})}\big) \right\vert ^{2}}{%
\left\vert v_{0}\sqrt{1-\frac{\omega _{L}^{2}}{F_{0}^{2}}(k_{z}-v_{0})}%
\right\vert },  \label{Gamma}
\end{equation}%
where $d_{z}(v)$ was defined in Eq. (\ref{dipole}).

With a bit of algebra Eq. (\ref{interf}) can be written as 
\begin{equation}
\sum_{j=1}^{N}\,\sum_{\alpha =1}^{2}e^{iS(t^{(j,\alpha
)})}=2\sum_{j=1}^{N}e^{i\bar{S}_{j}}\cos \left( \frac{\Delta S_{j}}{2}%
\right) ,\,  \label{sum}
\end{equation}%
where $\bar{S}_{j}=$ $\left[ S(t^{(j,1)})+S(t^{(j,2)})\right] /2$ is the
average action of the two trajectories released in the $j$th cycle, and $%
\Delta S_{j}=S(t^{(j,1)})-S(t^{(j,2)})$ is the accumulated action between
the two release times $t^{(j,1)}$ and $t^{(j,2)}$ within the same $j$-th
cycle. The two solutions of Eq. (\ref{ioniztimes+}) [Eq. (\ref{ioniztimes-}%
)] per optical cycle: The early release time $t^{(j,1)}$ and the late
release time $t^{(j,2)}$ lays within the first (or third) quarter of the $j$%
-th cycle and within the second (or fourth) quarter of the $j$-th cycle,
respectively.

From Eq. (\ref{action}), the semi-classical action along one electron
trajectory with ionization time $t^{(j,\alpha )}$ is, up to a constant,%
\begin{equation}
S(t^{(j,\alpha )})=\left( \frac{k_{z}^{2}}{2}+I_{p}+\frac{F_{L0}^{2}}{%
4\omega _{L}^{2}}-\omega _{X}\right) t^{(j,\alpha )}+\frac{F_{L0}}{\omega
_{L}^{2}}k_{z}\cos (\omega t^{(j,\alpha )})-\frac{F_{L0}^{2}}{8\omega
_{L}^{3}}\sin (2\omega t^{(j,\alpha )}).
\end{equation}%
The average action depends linearly on the cycle number $j$, 
\begin{equation}
\bar{S}_{j}=S_{0}+j\tilde{S},  \label{Savg}
\end{equation}%
where $S_{0}$ is a constant which will be cancelled out when taken the
absolute value in Eq. (\ref{interf}), and $\tilde{S}=\left( 2\pi /\omega
_{L}\right) \left( E+U_{p}+I_{p}-\omega _{X}\right) $.

On the other hand, the difference of the action $\Delta S_{j}$ is a constant
independent of the cycle number $j$, which can be expressed (dropping out
the subindex $j$) as%
\begin{eqnarray}
\Delta S &=&\left( \frac{k_{z}^{2}}{2}+I_{p}+U_{p}-\omega _{X}\right) \frac{1%
}{\omega _{L}}\left\{ \pi -2\sin ^{-1}\left[ \frac{\omega _{L}}{F_{L0}}%
\left\vert k_{z}-v_{0}\right\vert \right] \right\}  \nonumber \\
&&-\mathrm{sgn}(k_{z}-v_{0})\frac{F_{L0}}{2\omega _{L}^{2}}\left(
3k_{z}+v_{0}\right) \sqrt{1-\frac{\omega _{L}^{2}}{F_{L0}^{2}}\left(
k_{z}-v_{0}\right) ^{2}},  \label{DS}
\end{eqnarray}%
where $\mathrm{sgn}$ denotes the sign function. We note there is a discontinuity of $%
\Delta S$ for $k_{z}=v_{0}$. This occurs in the present case where the XUV pulse starts at the 
same time that $A_L =0$. In general, the discontinuity of $\Delta S$ depends on the delay between both pulses.
In the next section we show how this
discontinuity mirrors on the electron emission spectra.

In the same way (after some algebra) as for the case of ionization by a
monochromatic pulse \cite{Arbo08b,Arbo10a,Arbo10b,Arbo12}, Eq. (\ref{interf}%
) together with equations (\ref{sum}) and (\ref{Savg}) can be rewritten as 
\begin{equation}
\left\vert T_{if}\right\vert ^{2}=4\,\Gamma (k_{z})\underbrace{\cos
^{2}\left( \frac{\Delta S}{2}\right) }_{F(k_{z})}\underbrace{\left[ \frac{%
\sin \left( N\tilde{S}/2\right) }{\sin \left( \tilde{S}/2\right) }\right]
^{2}}_{B(k_{z})}.  \label{split}
\end{equation}%
Eq. (\ref{split}) indicates that the interference pattern can be factorized
in two contributions: (i) the interference stemming from a pair of
trajectories within the same cycle (\textit{intracycle} interference),
governed by the factor $F(k_{z}),$ and (ii) the interference stemming from
trajectories released at different cycles (\textit{intercycle} interference)
resulting in the well-known side bands (SBs) given by the factor $B(k_{z})$.
When $N\rightarrow \infty ,$ the second factor becomes a series of delta
functions, i.e., $B(k_{z})\rightarrow \sum_{n}\delta (E-E_{n}),$ where 
\begin{equation}
E_{n}=n\omega _{L}+\omega _{X}-I_{p}-U_{p}  \label{En}
\end{equation}%
are the positions of the SBs for the absorption of $n$ IR photons and one
XUV photon. When $n<0$ the emission of $\left\vert n\right\vert $ IR photons is meant,
whereas when $n=0,$ the ATI peak for the absorption of only one XUV
photon of frequency $\omega _{X}$ is described. It is worth to
notice that the energy of this ATI peak and the side bands are shifted
with the ponderomotive energy of the IR laser $U_{p}$ according to Eq. (\ref{En}).
The intracycle interference arises from the superposition of pairs of classical
trajectories separated by a time slit $\Delta t=t^{(j,2)}-t^{(j,1)}$ of the
order of less than half a period of the IR laser pulse, i.e., $\Delta t<\pi
/\omega _{L},$ giving access to emission time resolution of $\lesssim 1$ fs
(for near IR pulses), while the difference between $t^{(j,\alpha )}$ and $%
t^{(j+1,\alpha )}$ is $2\pi /\omega _{L}$, i.e., the optical period of the
IR laser. Equation (\ref{split}) is structurally equivalent to the intensity
for crystal diffraction: The factor $F(k_{z})$ represents the form (or
structure) factor accounting for interference modulations due to the
internal structure within the unit cell while the factor $B(k_{z})$ gives
rise to Bragg peaks due to the periodicity of the crystals.
The number $N$ of slits is determined by the duration of the XUV pulse
$\tau _{X}=2N\pi /\omega _{L}.$Therefore, $B(k_{z})$ in Eq.\ (\ref{split})
may be viewed as a diffraction grating in the time domain consisting of 
$N$ slits with $F(k_{z})$ being the diffraction factor for each slit.
As in each optical cycle there are two interfering electron trajectories,
it is reasonable to obtain a young-type intracycle interference pattern
of the form $F(k_z)=\cos^2 (\Delta S /2)$



\section{Results and discussion}
\label{results}

In order to compare the different methods described in the last section and
probe the general conclusion of the SCM that the momentum distribution can
be thought as the interplay between the inter- and intracycle interference
processes, we consider flattop envelopes for both the IR and XUV pulses. In
this sense, the IR laser field can be written as%
\begin{equation}
\vec{F}_{L}(t)=F_{L0}(t)\ \cos \left[ \omega _{L}\left( t-\frac{\tau _{L}}{2}%
\right) \right] \ \hat{z},
\label{IR-field}
\end{equation}%
where the envelope is given by%
\begin{equation}
F_{L0}(t)=F_{L0}\left\{ 
\begin{array}{ccc}
\frac{\omega _{L}t}{2\pi } & \mathrm{if} & 0\leq t\leq \frac{2\pi }{\omega
_{L}} \\ 
1 & \mathrm{if} & \frac{2\pi }{\omega _{L}}\leq t\leq \tau _{L}-\frac{2\pi }{%
\omega _{L}} \\ 
\frac{\left( \tau _{L}-t\right) \omega _{L}}{2\pi } & \mathrm{if} & \tau
_{L}-\frac{2\pi }{\omega _{L}}\leq t\leq \tau _{L}%
\end{array}%
\right.  \label{laser-envelope}
\end{equation}%
and zero otherwise so that the IR laser field is a cosine-like pulse
centered in the middle of the pulse, i.e., $t=\tau _{L}/2,$ where $\tau _{L}$ is the
laser pulse duration comprising an integer number of optical cycles with a
central flattop region and linear one-cycle ramp on and ramp off.

In the same way, we can define the XUV pulse as%
\begin{equation}
\vec{F}_{X}(t)=F_{X0}(t)\ \cos \left[ \omega _{X}\left( t-t_{12}-\frac{\tau
_{L}}{2}\right) \right] \ \hat{z},
\label{XUV-field}
\end{equation}%
where the main frequency of the XUV pulse is $\omega _{X}$ and we choose the
envelope as%
\begin{equation}
F_{X0}(t)=F_{X0}\left\{ 
\begin{array}{ccc}
\frac{\omega _{X}t}{2\pi } & \mathrm{if} & t_{b}\leq t\leq t_{b}+\frac{2\pi 
}{\omega _{X}} \\ 
1 & \mathrm{if} & t_{b}+\frac{2\pi }{\omega _{X}}\leq t\leq t_{e}-\frac{2\pi 
}{\omega _{X}} \\ 
\frac{\left( \tau _{X}-t\right) \omega _{X}}{2\pi } & \mathrm{if} & t_{e}-%
\frac{2\pi }{\omega _{X}}\leq t\leq t_{e}%
\end{array}%
\right. ,  \label{XUV-envelope}
\end{equation}%
and zero otherwise. We consider that there is an integer number
of optical cycles into the XUV pulse, i.e., $\tau _{X}/2\pi \omega _{X}$ is
integer, with linear one-cycle ramp on and ramp off. The time $t_{12}$
characterizes the delay between the centers of the IR and XUV pulses, and $%
t_{b}=t_{12}+\tau _{L}/2-\tau _{X}/2$ and $t_{e}=t_{12}+\tau _{L}/2+\tau _{X}/2$
denotes the beginning and the end of the XUV pulse, respectively.

In Fig. \ref{fields} (a) and (c) the total electric field $F(t)$ is plotted as a
function of time [as defined in equations (\ref{total-field}-\ref{XUV-envelope})]
with IR laser parameters $F_{L0}=0.02,$ $\omega _{L}=0.05,$ and $\tau _{L}=7T_{L}$,
and XUV pulse parameters $F_{X0}=0.01$ and $\omega _{X}=1.5$ with duration $\tau
_{X}=T_{L}$ in (a), and $\tau _{X}=3T_{L}$ in (c). The XUV pulse opens an
active window in the time domain for laser assisted XUV ionization marked
with a yellow shadow in Fig. \ref{fields}. The definitions of the IR and XUV pulses are
not capricious but they assure a flattop vector potential $A(t)$ fulfilling
the boundary conditions $A(0)=A(\tau _{L})=0.$ In Fig. \ref{fields} (b) and (d) we show
the values of $A(t)$ when $\tau _{X}=T_{L}$ and $\tau _{X}=3T_{L}$, respectively.
They look quite the same since for the chosen parameters the
amplitude of the vector potential of the IR laser pulse is $F_{L0}\omega
_{X}/F_{X0}\omega _{L}=60$ times higher than the amplitude of the XUV vector
potential.

\begin{figure}[tbp]
\includegraphics[width=0.5\textwidth]{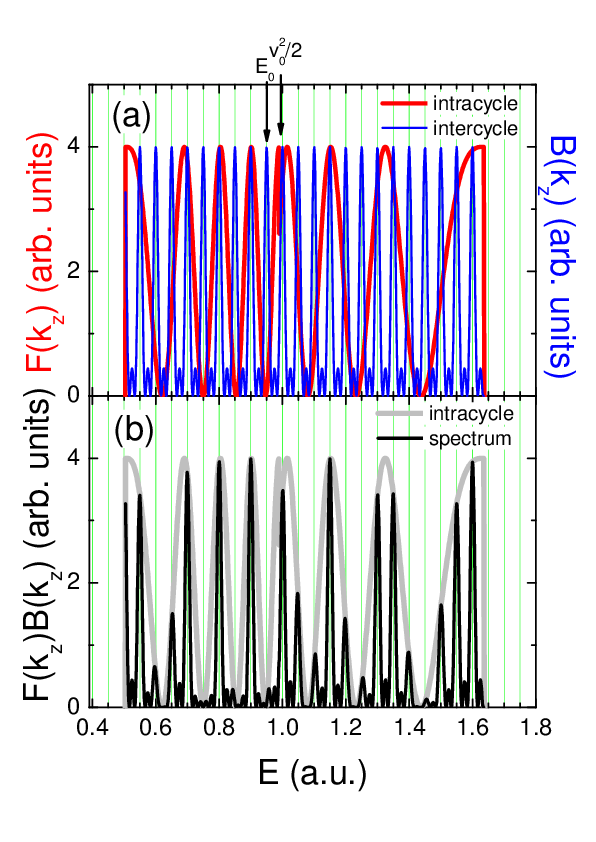}
\caption{(Color online) (a) Bildup of the interference pattern
following the SCM: Intracycle pattern given by the structure pattern $%
F(k_{z})$ in red thick line and intercycle interference given by the
function $B(k_{z})$ with $N=3$ [Eq. (\ref{split})]. (b) Total interference
pattern $F(k_{z})B(k_{z})$ with $N=3.$ The IR laser parameters are $%
F_{L0}=0.02,$ $\omega _{L}=0.05$, the XUV frequency $\omega _{X}=1.5$,
and delay time $t_{12}=0.$ Both fields are cosine-like. For the sake of
comparison, in light gray the intracycle pattern $F(k_{z})$ of (a). Vertical
lines depict the positions of the SBs $E_{n}$ of Eq. (\ref{En}).}
\label{SCM}
\end{figure}

In the following we analyze how the intercycle interference factor $B(k_{z})$
and the intracycle interference factor $F(k_{z})$ in Eq. (\ref{split})
within the SCM control the electron spectrum for LAPE. The factor $B(k_{z})$
calculated with the electric field described in Fig. \ref{fields} (a) is shown
in Fig. \ref{SCM} (a) in blue thin line as equispaced peaks with separation between
consecutive peaks equal to the IR laser frequency $\hbar \omega _{L}=0.05.$
The peaks of the function $B(k_{z})$ agree perfectly with the energies 
$E_{n} $ corresponding to the SBs [see Eq. (\ref{En})] marked with thin
vertical lines. For an arbitrary value of interfering optical cycles $N\geq 3$,
Eq. (\ref{split}) predicts $N-2$ secondary peaks  per optical cycle produced by the
interference of $N$ optical cycles (slits) in the laser pulse (diffraction
grating). In our case, two minima and a secondary peak is observed due to
the interference of three optical cycles ($\tau _{x}=3T_{L}$). The
intracycle structure factor $F(k_{z})$ shown in thick red curve displays
oscillations with maxima unrelated to the SBs. The positions of these maxima
can be calculated with $\Delta S=2m\pi ,$ with integer $m.$ The separation
of consecutive maxima of the intracycle factor $F(k_{z})$ depends on energy
in a nontrivial way. In this case, the separation of the consecutive
intracycle maxima is higher close to the classical boundaries $E=0.51$ a.u. and $E=1.65
$ a.u. than at intermediate energies. There is a discontinuity of the difference
of the action as a function of energy (and $k_{z}$) at $E=v_{0}^{2}/2$.
According to Eq. (\ref{saddle}), ionization times are calculated as the intersection
of the horizontal line $v_{0}-k_{z}$ with the vector potential $A(t)$.
When $k_{z}>v_{0},$ the two
ionization times lay in the second half of the optical cycle of the active
window [see Fig. \ref{fields} (b)]. As $k_{z}$ approaches $v_{0},$ the early release
time $t^{(1,1)}$ goes to the middle of the active window whereas the late
release times $t^{(1,2)}$ goes to the end of it. In turn, when $k_{z}<v_{0},$
the situation is different: As $k_{z}$ approaches $v_{0},$ the early release
time $t^{(1,1)}$ goes to the beginning of the active window whereas the late
release times $t^{(1,2)}$ goes to the middle. Such discontinuity does not
exist in the case of intracycle interference in above threshold
ionization by an IR pulse since, in that case, $v_{0}=0$ (there is no XUV
pulse) in Eq. (\ref{DS}) \cite{Arbo08b,Arbo10a,Arbo10b,Arbo12}.

In Fig. \ref{SCM} (b) we show the interference pattern for the case of $N=3$
interfering cycles into the active window [Fig. \ref{fields} (c) and (d)]. Only the
factor $B(k_{z})F(k_{z})$ is displayed setting the variation of the
ionization rate $\Gamma (k_{z})$ to unity to focus on the interference
process. For the sake of comparison, in light gray the intracycle pattern $%
F(k_{z})$ of (a) is also displayed. We observe that the intercycle SB peaks
given by $B(k_{z})$ [Fig. \ref{SCM} (a)] are modulated by the intracycle
interference factor $F(k_{z})$. The intracycle
interference can lead to the suppression of SBs (for example, near $E=1.43$).

\begin{figure}[tbp]
\includegraphics[width=0.5\textwidth]{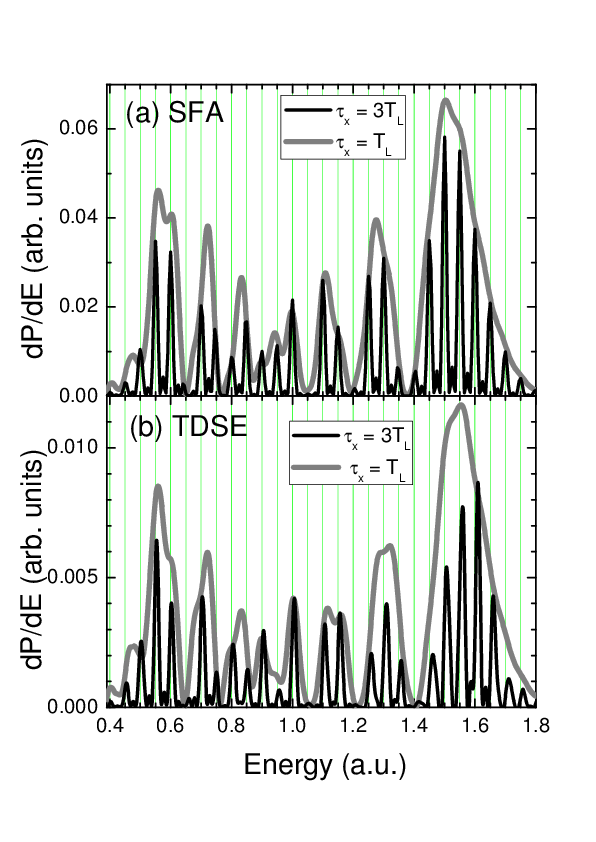}
\caption{(Color online) Energy distribution in the forward direction
for the same laser parameters and XUV frequency as in Fig. \ref{fields} calculated
within (a) the SFA and (b) the TDSE. The laser duration is $\tau
_{L}=7T_{L}=879.65$, and the XUV field amplitude $F_{X0}=0.01$. Vertical
lines depict the positions of the SBs $E_{n}$ of Eq. (\ref{En}).}
\label{Edis}
\end{figure}

We need to compare our SCM with quantum SFA and \textit{ab initio}
calculations by solving numerically the TDSE. In Fig. \ref{Edis} we plot the energy
distribution of electron emission in the forward direction for the same
laser pulse described in Fig. \ref{fields}. We perform SFA and TDSE calculations in
Fig. \ref{Edis} (a) and Fig. \ref{Edis} (b), respectively, for two different durations of the
XUV pulse. For the case of $\tau _{X}=3T_{L}$ we observe a set of peaks
separated by the laser frequency $\omega _{L}$ in agreement with Eq. (\ref%
{En}) whose positions are illustrated with vertical thin lines. By comparing
Fig. \ref{SCM} and Fig. \ref{Edis}, we see that the quantum (TDSE and SFA) energy
distributions extend about $0.2$ a.u. beyond the classical limits
$\left( v_{0}\mp F_{L0}/\omega _{L}\right) ^{2}/2$. The
agreement between SFA and TDSE results is remarkable and both are
qualitatively similar to the SCM of Fig. \ref{SCM}. We would like to point out that
the energy distributions exhibit sharp modulations in agreement with the
intracycle interference pattern calculated with an XUV pulse duration $\tau
_{X}=T_{L}$ in gray thick line. In this sense, the fact that the intracycle
interference pattern modulates the sidebands in the energy distribution,
albeit derived within the SCM, is also valid for the quantum calculations.
The reason for this is under current investigation, however we note that 
it is in close relationship with previous work \cite{Kazansky10,Bivona10} where
the PE spectra is factorized as two contributions.
It is also worth to
mention that, as within the SCM, there are frustrated SBs, i.e., close to 
$E=0.63,$ $1.03,$ $1.2,$ and $1.4$ coinciding with the minima of the intracycle
interference pattern

\begin{figure}[tbp]
\includegraphics[width=0.5\textwidth]{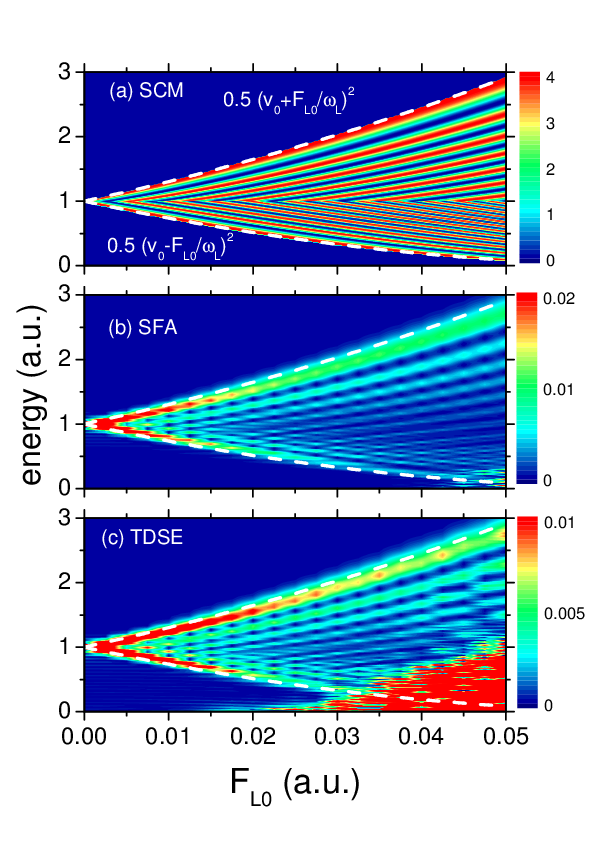}
\caption{(Color online) Photoelectron spectra in the forward
direction (in arbitrary units) calculated at different laser field strengths
within (a) the SCM, (b) the SFA, and (c) the TDSE. The IR laser frequency is 
$\omega _{L}=0.05$ and the XUV pulse have $F_{X0}=0.01$, $\omega _{X}=1.5$,
and $\tau _{X}=T_{L}.$ Both fields are cosine-like. In dashed line we show
the classical boundaries of LAPE. The high intensity \textquotedblleft
spot\textquotedblright\ in the right bottom corner of (c) corresponds to
ionization by the IR laser pulse alone.}
\label{EFL0}
\end{figure}

In order to investigate the dependence of the intracycle interference
pattern on the intensity of the laser pulse, we perform calculations of the
energy distribution in the forward direction within the SCM in Fig. \ref{EFL0} (a),
the SFA in Fig. \ref{EFL0} (b), and the TDSE in Fig. \ref{EFL0} (c) for laser field amplitudes
from $F_{L0}=0$ up to $0.05.$ In this sense, the intracycle pattern in Fig. \ref{SCM} (a) is a cut of Fig. \ref{EFL0} (a) at $F_{L0}=0.02.$ The same applies to the
intracycle patterns of Fig. \ref{Edis} (a) and (b) with Fig. \ref{EFL0} (b) and (c),
respectively. The classical boundaries $\left( v_{0}\mp F_{L0}/\omega
_{L}\right) ^{2}/2$ are drawn in dash lines and they exactly delimit the SCM
spectrogram of Fig. \ref{EFL0} (a), as expected. The discontinuity at 
$E=v_{0}^{2}/2=1 $ is clearly independent of the laser field amplitude.
Above the discontinuity, the interference maxima (and minima) exhibit a
positive slope as a function of $F_{L0},$ whereas below it, the stripes have
negative slope. The classical boundaries slightly blur for the SFA
spectrogram, also showing the characteristic intracycle stripes with
positive (negative) slope close to the top (bottom) classical boundary. For
intermediate energies close to $E\simeq 1,$ it is difficult to determine
such discontinuity. In Fig. \ref{EFL0} (c), the TDSE calculation exhibit a strong
probability distribution for high values of $F_{L0}$ in the low energy
region. The source of this enhancement of the probability is atomic ionization by
the IR laser pulse alone, which has not been considered in our SCM and is strongly
suppressed in the SFA because the laser photon energy is much less than the
ionization potential, i.e., $\omega _{L}\ll I_{p}.$ For this reason we can
confirm that the SFA is a very reliable method to deal with LAPE rather than
ATI by IR lasers. Except for the region where ionization by the laser field alone
becomes important, SFA and TDSE spectrograms agree with each other and
qualitatively resemble the SCM calculations.

\begin{figure}[tbp]
\includegraphics[width=0.5\textwidth]{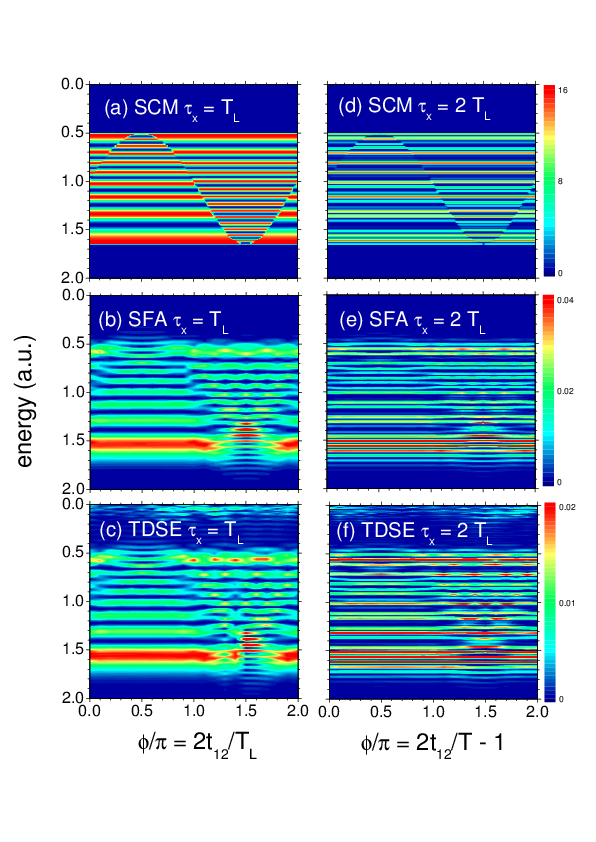}
\caption{(Color online) Photoelectron spectra in the forward
direction (in arbitrary units) as a function of the time delay $t_{12}$
within (a,d) the SCM, (b,e) the SFA, and (c,d) the TDSE. The IR laser
frequency is $\omega _{L}=0.05$ and $\tau _{L}=7T_{L}$ and the XUV pulse
have $F_{X0}=0.01$, $\omega _{X}=1.5$, XUV pulse duration $\tau _{X}=T_{L}$
in (a), (b), and (c), and $\tau _{X}=2T_{L}$ in (d), (e), and (f). Both fields are cosine-like.}
\label{Et12}
\end{figure}

So far, we have performed our analysis of the electron emission in the
forward direction for zero time delay, i.e., the center of the IR laser and
XUV pulses coincide as $t_{12}=0.$ In order to reveal how the intracycle
interference pattern changes with the time delay, we vary $t_{12}$ from $0$
up to $T_{L}.$ This means that the center of the XUV pulse situated at 
$t_{12}$ corresponds to a phase into the laser optical cycle 
$\phi =\omega_{L}t_{12}=2\pi t_{12}/T_{L}.$ In Fig. \ref{Et12} (a) we show
the SCM intracycle interference pattern in the forward direction as a function
of the time delay $t_{12}$. The horizontal stripes show the independence of the
intracycle interference pattern with the time delay, except for the
discontinuity in Eq. (\ref{DS}) for values of energy equal to 
$E_{\mathrm{disc}}=\left[ v_{0}+A(\tau_L+t_{12}-T_{L}/2)\right] ^{2}/2$.
For $t_{12}=0,$ the discontinuity is situated at $E_{\mathrm{disc}}=v_{0}^{2}/2$
since $A(\tau_L-T_{L}/2)=0$ as is shown in Fig. \ref{SCM}. As $t_{12}$ 
(and $\phi $) varies, the discontinuity follows the shape of the vector potential.
For the cases that $\phi =\pi /2$ and $3\pi /2$ the discontinuity moves to the
classical boundary loosing entity. For the case $\phi =\pi /2,$ the
separation between consecutive intracycle interference stripes is smaller
for lower energies and increases as the energy grows. Contrarily, for $\phi
=3\pi /2,$ energy separation between consecutive intracycle interference
maxima is higher for lower energy and diminishes as energy increases. The
SFA and TDSE energy distribution in Fig. \ref{Et12} (b) and (c), respectively,
exhibit similar characteristics to the SCM. Interestingly, the discontinuity
at $E_{\mathrm{disc}}$ can be clearly observed for the same energy values.
The resemblance between the SFA and TDSE results is remarkable, which shows
that the SFA is a very appropriate method to deal with LAPE processes and
computationally much less demanding than solving the TDSE \textit{ab initio}.
Low energy contributions in TDSE calculations [Fig. \ref{Et12} (c)] are due to
ionization by the interaction between the atom and the IR laser pulse alone.
There are two characteristics of SFA and TDSE spectra which deserve more
study: (i) For $\phi \simeq \pi /2$ the energy distribution extends to lower
energy values than for other $\phi $ values $(E\simeq 0.5)$, whereas for $%
\phi \simeq 3\pi /2$ it extends for higher energy values ($E\simeq 1.7$),
and (ii) the horizontal intracycle interference stripes show some structure
at the right of the above mentioned discontinuity, i.e., $E<E_{\mathrm{disc}%
} $ which is absent at the left of it, i.e., $E>E_{\mathrm{disc}}$.

When we calculate the energy distribution for a XUV pulse with duration 
$\tau _{X}=2T_{L}$, our active window comprises two IR optical cycles. For
the particular case of zero time delay, i.e., $t_{12}=0$ (both IR and XUV
pulses centered at the same instant of time), the vector potential at the
beginning of the active window has a change of sign compared to the $\tau
_{X}=T_{L}$ case. Therefore, for the sake of comparison, we redefine the
phase $\phi =\omega _{L}t_{12}-\pi =2\pi t_{12}/T_{L}-\pi ,$ varying the
time delay $t_{12}$ from $T_{L}/2$ up to $3T_{L}/2.$ In this sense, with
this new definition, $\phi =0$ corresponds to $t_{12}=T_{L}/2,$ with the
same behavior of the vector potential inside the active window. In Fig. \ref{Et12}
(d) the SCM spectrum display horizontal lines corresponding to the
intercycle interference, which are modulated by the intracycle pattern of
Fig. \ref{Et12} (a). The discontinuity of the intracycle modulation can also be
observed, which stands for the SFA [Fig. \ref{Et12} (e)] and TDSE [Fig. \ref{Et12} (f)]
too. Once again, the agreement between the SFA and TDSE is very good, with the
exception of the low energy contribution due to the ionization by the IR
laser pulse in the TDSE spectrogram. By comparing the intracycle pattern for 
$\tau_X = T_L$ on the left column [Figs. \ref{Et12} (a), (b), and (c)]
to the whole interference pattern for $\tau_X = 2 T_L$ on the right column
[Figs. \ref{Et12} (d), (e), and (f)], we see the interplay between intra-
and intercycle interference, i.e., the intracycle interference pattern
works as a modulation of the intercycle interference pattern (SBs)
for the active window with duration of two optical laser cycles.

\begin{figure}[tbp]
\includegraphics[width=0.5\textwidth]{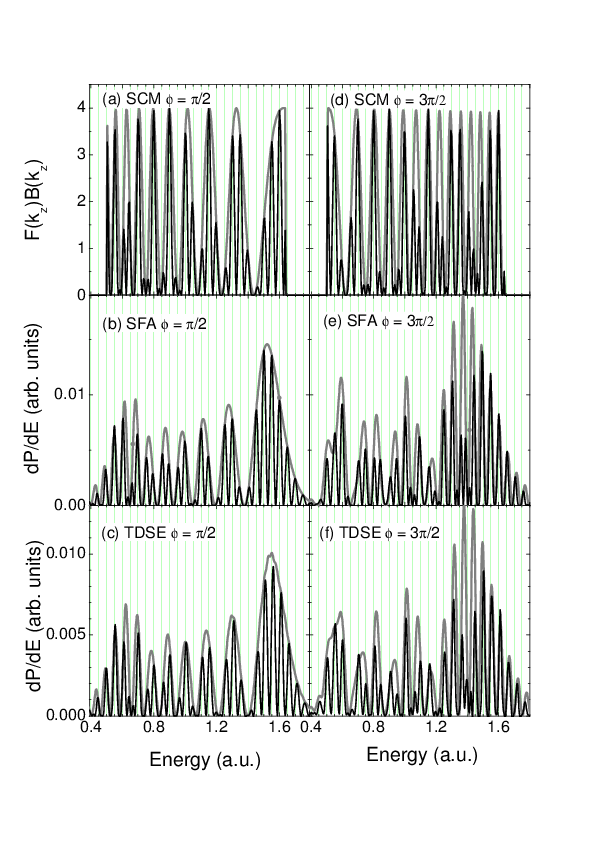}
\caption{(Color online) Photoelectron spectra in the forward
direction (in arbitrary units) for time delays $t_{12}$ corresponding to
(a), (b), and (c) $\phi =\pi /2,$ and (d), (e), and (f) $\phi =3\pi /2$,
within (a,d) the SCM, (b,e) the SFA, and (c,d) the TDSE. The IR laser
frequency is $\omega _{L}=0.05$ and $\tau _{L}=7T_{L}$ and the XUV pulse
parameters are $F_{X0}=0.01$, $\omega _{X}=1.5$, and $\tau _{X}=T_{L}$
(thick light grey curve) and $\tau _{X}=2T_{L}$ (thick light grey curve). Both fields are cosine-like.}
\label{Et12cases}
\end{figure}

The intracycle energy distributions ($\tau _{X}=T_{L}$) in Fig. \ref{SCM}, Fig. \ref{Edis}
(a), and Fig. \ref{Edis} (b) can be regarded as cuts of the intracycle interferograms
of Figs. \ref{Et12} (a), (b), and (c), respectively. For the sake of completeness, we
show also in the left column of Fig. \ref{Et12cases} [(a), (b), and (c)] the energy
distribution for $\phi =\pi /2$ for $\tau _{X}=T_{L}$ and $\tau _{X}=2T_{L}.$
We observe how in all calculations [SCM in Fig. \ref{Et12cases} (a), SFA in Fig. \ref{Et12cases} (b),
and TDSE in Fig. \ref{Et12cases} (c)] the separation of consecutive intracycle maximum
grows as the energy increases. The energy distributions for $\tau
_{X}=2T_{L} $ show a SB structure (intercycle interference) modulated by the
intracycle interference pattern. As the separation of intracycle maxima of
the intracycle interference pattern near the lower classical limit is close
to the laser photon energy $\hbar \omega _{L}$, it competes with the
intercycle interference pattern (SBs) for $\tau _{X}=2T_{L}$ whose
separation is also $\hbar \omega _{L}.$ Therefore, the interplay of intra-
and intercycle interference pattern gives rise to new oscillation structures
of the energy distribution by ionization of the XUV pulse of duration $\tau
_{X}=2T_{L}$ assisted by the laser pulse. For example, a gross structure is
observed Fig. \ref{Et12cases} (b) and (c) with a minimum at $E\simeq 0.63$ for the SFA and
TDSE. The same is valid for the phase $\phi =3\pi /2$ in Figs. \ref{Et12cases} (d), (e),
and (f) for the SCM, SFA, and TDSE, respectively. However, in this case, the
competition between intra- and intercycle interference patterns takes place
close to the higher classical limit. Again, a grosser structure is formed
making the energy distributions when $\tau _{X}=2T_{L}$ for $\phi =\pi /2$
and $\phi =3\pi /2$ to be similar. It is expected that as the active
window gets wider, i.e., $\tau _{X}\gg T_{L},$ the agreement between energy
spectra for $\phi =\pi /2$ and $\phi =3\pi /2$ improves.

\begin{figure}[tbp]
\includegraphics[width=0.5\textwidth]{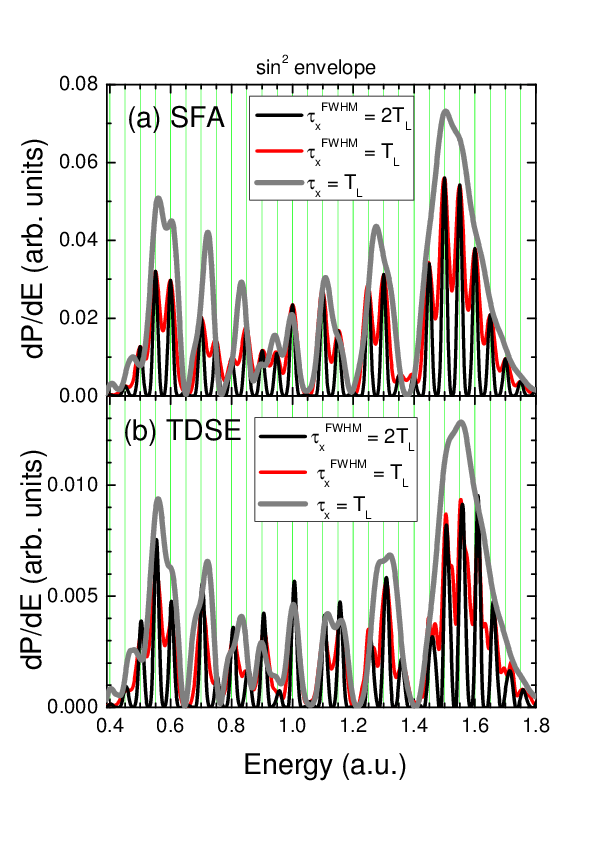}
\caption{(Color online) Energy distribution in the
forward direction calculated within (a) the SFA and (b) the TDSE. The laser
duration is $\tau _{L}=9T_{L}=1130.97$ and the envelope function of the XUV
pulse is $F_{X0}(t)=\sin ^{2}(\pi t/\tau _{X})$ with $\tau _{X}=345.2$ and $%
690.4$ which correspond to a FWHM duration of $\tau _{X}^{\mathrm{FWHM}%
}=T_{L}$ and $2T_{L}$, respectively. For completeness, we also show the
intracycle momentum for the flattop XUV pulse of Fig. \ref{Edis}. Vertical lines
depict the positions of the SBs $E_{n}$ of Eq. (\ref{En}).}
\label{sin2}
\end{figure}

In the following we investigate the role of the envelope of the XUV pulse $%
F_{X0}(t)$ on LAPE. So far, we have used the trapezoidal envelope given by
Eq. (\ref{XUV-envelope}), which opens a well-defined active window of
duration almost equal to $\tau _{X}$ (since it has a one-cycle ramp on and
one-cycle ramp off of duration $T_{X}=2\pi /\omega _{X}\ll T_{L}$ each).
Now, we consider an XUV pulse with squared sine envelope%
\begin{equation}
F_{X0}(t)=\sin ^{2}\left( \frac{\pi t}{\tau _{X}}\right) =\sin ^{2}\left( 
\frac{\pi t}{2\tau _{X}^{\mathrm{FWHM}}}\right) ,  \label{sin2-env}
\end{equation}%
where $\tau _{X}^{\mathrm{FWHM}}=\tau _{X}/2$ is the FWHM duration of the
electric field. The result for the energy distribution in the forward
direction due to an XUV pulse with $\sin ^{2}$ envelope with 
$\tau _{X}^{\mathrm{FWHM}}=T_{L}$ assisted by the laser pulse described in equations 
(\ref{IR-field}) and (\ref{laser-envelope}) shows modulated SB peaks Fig. \ref{sin2}
(a) and (b) calculated within the SFA and TDSE, respectively. When we
compare these results with the energy distribution calculated with the
flattop pulse of equations (\ref{XUV-field}) and (\ref{XUV-envelope}), we
realize that the origin of the modulations of the SBs are due to the
intracycle interference also for the $\sin^{2}$ XUV envelope of Eq. (\ref{sin2-env}).
When we double the duration of the XUV pulse, i.e., 
$\tau _{X}^{\mathrm{FWHM}}=2T_{L}$, the SB peaks are sharper
because the double of the optical cycles are involved in the active window enhancing,
consequently, the intercycle interference giving rise to almost perfect destructive
interference (minima of the energy distribution are zero).
The agreement between SFA and TDSE calculations is very good.
We can say, therefore, that the envelope of the XUV pulse play a minor role
in LAPE and most of the conclusions derived for the flattop XUV pulse
are still valid for a smooth experimental-like envelope shape.


\section{\label{conc}Conclusions}


We have studied the electron emission in the forward direction produced by hydrogen
ionization subject to a XUV and laser pulse both linearly polarized in the same direction.
The PE spectrum 
can be regarded as an interference pattern
of a diffraction grating in the time domain. Semiclassically, the intercycle interference of
electron trajectories from different optical cycles gives rise to side bands, whereas
the intracycle interference of electron trajectories born in the same optical cycle
originates a coarse grained pattern modulating the side bands. 
We have observed 
that the SFA is sufficiently accurate to describe the photoelectron spectrum
when compared to the TDSE. The intracycle pattern is independent of the
XUV pulse duration and envelope but exhibits a jump at a given energy as a function of the
time delay between the two pulses $t_{12}$ reproducing the profile of the laser vector
potential. 

\begin{acknowledgments}
Work supported by CONICET PIP0386, PICT-2012-3004 and PICT-2014-2363 of ANPCyT (Argentina), and the University of Buenos Aires (UBACyT 20020130100617BA). Work triggered by discussion between
Pablo Macri and D.G.A.
\end{acknowledgments}

\bibliography{bibliography}
\nocite{*}

\end{document}